\documentclass[sigconf]{acmart}
\usepackage{bm}
\AtBeginDocument{%
  \providecommand\BibTeX{{%
    \normalfont B\kern-0.5em{\scshape i\kern-0.25em b}\kern-0.8em\TeX}}}

\setcopyright{acmlicensed}
\copyrightyear{2018}
\acmYear{2018}
\acmDOI{XXXXXXX.XXXXXXX}

\acmConference[Conference acronym 'XX]{Make sure to enter the correct
  conference title from your rights confirmation emai}{June 03--05,
  2018}{Woodstock, NY}
\acmISBN{978-1-4503-XXXX-X/18/06}

\begin{document}

\title{Investigating Role of Big Five Personality Traits \\ in Audio-Visual Rapport Estimation}
\author{Takato Hayashi}
\email{hayashi0884@jaist.ac.jp}
\author{Ryusei Kimura}
\author{Shogo Okada}
\email{ryusei_kimura@jaist.ac.jp}
\affiliation{%
  \institution{Japan Advanced Institute of Science and Technology}
  \city{Nomi}
  \state{Ishikawa}
  \country{Japan}
}

\author{Ryo Ishii}
\affiliation{%
  \institution{Human Informatics Laboratories, NTT Corporation}
  \city{Yokosuka}
  \state{Kanagawa}
  \country{Japan}}

\renewcommand{\shortauthors}{Hayashi et al.}

\begin{abstract}
Automatic rapport estimation in social interactions is a central component of affective computing. Recent reports have shown that the estimation performance of rapport in initial interactions can be improved by using the participant's personality traits as the model's input. In this study, we investigate whether this findings applies to interactions between friends by developing rapport estimation models that utilize nonverbal cues (audio and facial expressions) as inputs. Our experimental results show that adding Big Five features (BFFs) to nonverbal features can improve the estimation performance of self-reported rapport in dyadic interactions between friends. Next, we demystify how BFFs improve the estimation performance of rapport through a comparative analysis between models with and without BFFs. We decompose rapport ratings into perceiver effects (people's tendency to rate other people), target effects (people's tendency to be rated by other people), and relationship effects (people's unique ratings for a specific person) using the social relations model. We then analyze the extent to which BFFs contribute to capturing each effect. Our analysis demonstrates that the perceiver’s and the target's BFFs lead estimation models to capture the perceiver and the target effects, respectively. Furthermore, our experimental results indicate that the combinations of facial expression features and BFFs achieve best estimation performances not only in estimating rapport ratings, but also in estimating three effects. Our study is the first step toward understanding why personality-aware estimation models of interpersonal perception accomplish high estimation performance. 
\footnote{Code will be available upon acceptance.}
\end{abstract}




\maketitle

\section{Introduction}
\label{Introduction}
The term \textit{rapport} can be defined as the feeling of being “in sync” with a conversational partner \cite{Huang2011-qc}. If a machine learning model can estimate the rapport of dyads, the system will be able to provide various types of support according to the estimates. For example, the rapport between students leads to learning gains in peer tutoring \cite{Sinha2015-lg}, so if teachers have access to the estimates, they can support dyads (pairs of students) more selectively. In these cases, dyads are already in a close relationship to some extent. Previous research \cite{Cerekovic2017-tq, Hagad2011-od, Hayashi2023-bp, Madaio2017-xr, Muller2018-ax, Sharma2021-mi, Zhao2016-dk} has addressed rapport estimation in initial interactions, but despite its importance, only a few studies \cite{Madaio2017-xr, Zhao2016-dk} have addressed it in interactions between intimate participants. Thus, we focus here on rapport estimation in dyadic interactions between friends.

Our final goal is to clarify how to maximize the potential of rapport estimation models by using personality traits in interactions between friends. In general rapport estimation, the input of rapport estimation models is the participant's behavioral features in an interaction, and the output is a value or class indicating the degree of rapport. Among the previous studies on automatic rapport estimation \cite{Cerekovic2017-tq, Muller2018-ax}, a key finding is that the use of personality trait features based on the five-factor model can improve the estimation performances of rapport in initial interactions. In addition to the empirical evidence, various theoretical findings regarding personality traits and interpersonal perceptions (e.g., rapport) have been reported in the field of social psychology \cite{Cuperman2009-fp, Funder1993-xi, Harris2016-xv, Wood2010-sa}. For example, Cuperman et al. \cite{Cuperman2009-fp} showed that participants with high agreeableness tend to rate rapport toward their conversational partners highly.

\begin{figure}[t]
    \centering
    \includegraphics[width=0.9\linewidth]{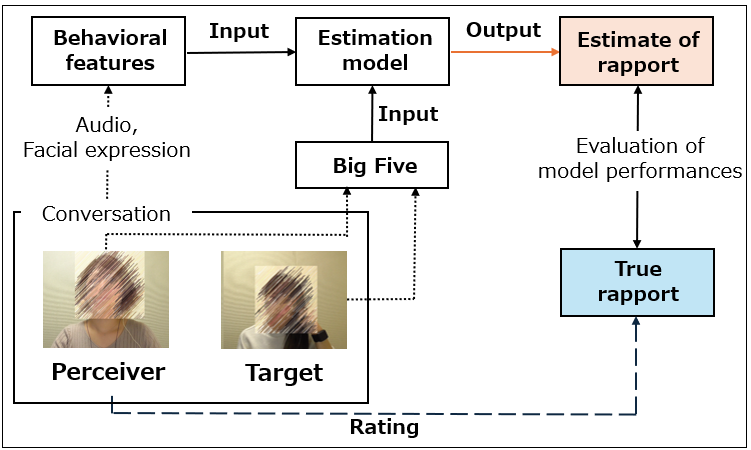}
    \caption{Overview of rapport estimation.}
    \label{fig:overview_rapport_est}
\end{figure}

\textbf{Our first research objective (RO-1)} is to clarify whether Big Five features (BFFs) can improve the estimation performance of rapport in interactions between friends. As stated above, it has already been established that adding personality trait features to behavioral features improves the estimation performance of rapport \cite{Cerekovic2017-tq, Muller2018-ax}. However, it is unclear whether this finding applies to interactions between friends since the association between personality traits and rapport changes depending on the relationship between participants. For example, according to Tickle-Degnen and Rosenthal, \textit{positivity} is important for building rapport in initial interactions, but its importance decreases as intimacy among participants increases \cite{Tickle-Degnen1990-jp}. Thus, even though extraversion (the Big Five dimension related to \textit{positivity}) is a strong cue for estimating rapport in initial interactions, it may be a weak cue in interactions between friends. 

To achieve RO-1, we develop personality-aware rapport estimation models (see Fig. \ref{fig:overview_rapport_est}). Furthermore, we investigate how to use BFFs to achieve higher performance improvement. To this end, we combine the BFFs with audio, facial expressions, and bimodal features, and then compare their performance. We also compare the performance under three conditions: 1) only the BFFs of a perceiver (person who rates rapport) are accessible, 2) only the BFFs of a target (person whose rapport is rated) are accessible, and 3) both the perceiver’s and the target’s BFFs are accessible. 

\textbf{The second research objective (RO-2)} is to reveal the role of BFFs in the performance improvement through comparisons between the model with and without BFFs. Our experimental results related to RO-1 show that BFFs substantially improve the rapport estimation performance. The ratings of interpersonal perception consist of three components: (a) people’s overall tendencies to rate other people (\textit{perceiver effect}), (b) people’s overall tendencies to be rated by other people (\textit{target effect}), and (c) people’s unique ratings for a specific partner beyond the perceiver and target effects (\textit{relationship effect}) \cite{Joel2017-ne, Kenny2019-ji}. These effects represent the different aspects of interpersonal perceptions. Thus, if the estimation performance of rapport ratings improves, the model becomes able to capture any or all of these effects more accurately. At the same time, better capturing each effect leads to better estimation performances of rapport ratings. Hence, we reveal the extent to which performance gains by BFFs can be attributed to capturing each effect.

To calculate three effects, we introduce an analysis approach based on a social relations model (SRM) \cite{Kenny2019-ji, Kenny1984-su}, which is a tool for understanding human perception between two individuals. First, we decompose true (human) ratings and the estimates of ratings into the three effects using SRM. Second, to measure how a model can capture these effects, we calculate the Pearson's product-moment correlation coefficient (PCC) and concordance correlation coefficient (CCC) between the true effects and the estimates of effects. Finally, we compare the PCC and CCC for each effect between the model with and without BFFs.

For RO-2, we come up with three hypotheses and verify them: 1) the perceiver’s BFFs lead models to capture perceiver effects, 2) the target’s BFFs lead models to capture target effects, and 3) the combinations of both BFFs lead models to capture relationship effects. The first and second hypotheses are based on evidence that the perceiver's Big Five personality traits influence the perceiver's tendency to rate rapport with other people and evidence that the target's Big Five personality traits influence the target's tendency to be rated by other people, respectively \cite{Cuperman2009-fp}. The third hypothesis is based on evidence that the specific combinations of the Big Five personality traits (e.g., both have high agreeableness) lead to uniquely high or low rapport \cite{Cuperman2009-fp}.

The motivation behind the two research objectives is to reveal findings that help us develop a better rapport estimation model among intimate dyads. RO-1 provides readers with information to judge whether performance gain is worth the cost of collecting Big Five personality traits. As for achieving RO-2, it provides interpretability about the performance gain by BFFs and clarifies which types of BFFs are suitable for obtaining the desired auxiliary information. If the models can capture any or all three effects, an estimate of the rapport ratings supplies the auxiliary information. For example, estimates of target effects are useful for detecting exceptionally undesirable people (i.e., low target effect). Furthermore, beyond the limit of rapport and BFFs, our analysis approach is helpful in analyzing the extent to which model improvements or data extensions have contributed to capturing different aspects of interpersonal perception.

In summation, our contributions are delineated as follows:
\begin{itemize}
  \item Our experimental results show that BFFs can improve the estimation performance of self-reported rapport in dyadic interactions between friends.
  \item We introduce an analysis approach using SRM, which reveals the extent to which the model can capture perceiver, target, and relationship effects.
  \item We present evidence that the perceiver’s and target's BFFs lead estimation models to accurately capture the perceiver and target effects, respectively.
  \item We demonstrate that the combinations of facial expression and BFFs achieve best estimation performances not only in estimating rapport ratings, but also in estimating three effects.
\end{itemize}

Section \ref{Related_Works} provides an overview of related works. Section \ref{Data} introduces our dataset and analysis results, and Section \ref{Method} explains the methods of feature extraction, model, and analysis approach using SRM. In Section \ref{experimental_settings}, we describe the experimental settings. Section \ref{RESULT} shows our experimental results, and in Section \ref{Discussion}, we discuss their implications. 

\section{RELATED WORKS}
\label{Related_Works}
\subsection{Rapport and Machine Learning}
As rapport plays an essential role in building good relationships with others, it is an extensively researched topic in social psychology. Bernieri et al. \cite{Bernieri1996-it} outlined nonverbal cues indicating rapport in various contexts. Tickle-Degnen and Rosenthal \cite{Tickle-Degnen1990-jp} demonstrated that the key nonverbal cues indicating rapport can change depending on the relationship between participants. Harrigan et al. \cite{Harrigan1985-cd} found significant differences of eye movement and gesture between high- and low-rapport doctors. Furthermore, Miles et al. \cite{MILES2009585} showed that synchrony between two participants is associated with high rapport in both visual and audio cues. Grahe and Bernieri \cite{Grahe1999-tm} clarified that visual cues are the most useful for the observer to perceive rapport accurately. Overall, previous studies provide evidence that visual cues are more strongly associated with rapport than audio cues.

Inspired by findings on the relationships between nonverbal cues and rapport, recent researchers in machine learning have addressed automatic rapport estimation. Hayashi et al. \cite{Hayashi2024, Hayashi2023-bp} proposed a ranking model to rank conversation partners based on the degree of self-reported rapport. Other studies \cite{Madaio2017-xr, Zhao2016-dk} have estimated the rapport between two students in a peer tutoring scenario. As with studies in social psychology, visual features are found to be more useful cues for rapport estimation than audio features \cite{Cerekovic2017-tq, Hagad2011-od, Muller2018-ax, Wang2009-mh}.

Among the previous studies on rapport estimation, a key finding is that the use of personality trait features based on the five-factor model can improve the estimation performances in initial interactions \cite{Cerekovic2017-tq, Muller2018-ax}. Cerekovic et al. \cite{Cerekovic2017-tq} showed that a model using features combining social cues and personality traits achieves the best estimation performance of rapport for virtual agents. Similarly, in the task of detecting low rapport in the early stages of group interaction, Muller et al. \cite{Muller2018-ax} found that adding personality trait features to facial expression features improves the estimation performance. These two studies formulated rapport estimation as classifying the high/low rapport class. However, in these cases, the estimation results have only a little information about the degree of rapport. Furthermore, there is a risk of adding bias due to the class splitting criteria \cite{Martínez2014}. Therefore, in this study, we formulate rapport estimation as regression. In addition, in contrast to previous studies investigating initial interactions, we focus on the usefulness of personality traits in friend interactions.

\subsection{Interpersonal Perception and Social Relations Model}
The ratings of interpersonal perception consist of three components: perceiver, target, and relationship effects \cite{Kenny2019-ji}. These effects represent the different aspects of interpersonal perceptions. As an example, take the perception of rapport that John feels for Tom. The perception can be decomposed into three components: a) John’s tendencies to rate with other people (John’s perceiver effect), b) Tom’s tendencies to be rated by other people (Tom’s target effect), and c) John’s unique ratings for Tom beyond perceiver and target effect (John’s relationship effect to Tom).

These effects are typically calculated using a social relations model (SRM) \cite{Kenny2019-ji, Kenny1984-su}, which is a tool for analyzing human perception between two individuals. In this section, we explain how to calculate these effects with the settings of a round-robin design, in which the same participant serves as both a perceiver and a target (see table in Fig. \ref{fig:AnalyticalApproach}). According to SRM \cite{Kenny2019-ji, Kenny1984-su}, the rapport score by perceiver $i$ to target $j$ in group $k$ consists of the perceiver effect, the target effect, the relationship effect, and the group average:
\begin{equation}
X_{ijk} =  p_i + t_j + g_{ij} + M_{..k},
\end{equation}
where $p_i$ is the perceiver effect for person $i$; $t_j$ is the target effect for person $j$; $r_{ij}$ is the relationship effect for $i$ with $j$; and $M_{..k}$ is the mean of the ratings given by all participants in group $k$. The perceiver effect represents how the perceiver views other people on average, and the target effect represents how the target is viewed on average by others. The relationship effect represents the perceiver’s unique view toward the target. The perceiver effect ($p_i$) is
\begin{equation}
p_i = \frac{(n-1)^2}{n(n-2)} M_{i . k}+\frac{n-1}{n(n-2)} M_{. i k}-\frac{n-1}{n-2} M_{. . k},
\end{equation}
and the target effect ($t_i$) is calcurated as
\begin{equation}
t_i = \frac{(n-1)^2}{n(n-2)} M_{. i k}+\frac{n-1}{n(n-2)} M_{i . k}-\frac{n-1}{n-2} M_{. . k},
\end{equation}
where $M_{i.k}$ is the mean of the ratings given by participant $i$ in group $k$. $M_{.ik}$ is the mean of the ratings given to participant $i$ in group $k$, and $n$ is the group size. The relationship effect for perceiver $i$ with target $j$ ($g_{ij}$) is
\begin{equation}
g_{ij} = X_{ijk} - p_i - t_i - M_{..k}.
\end{equation}

Previous studies have reported theoretical findings regarding the relationship between personality traits and interpersonal perceptions \cite{Cuperman2009-fp, Funder1993-xi, Harris2016-xv, Wood2010-sa}. Cuperman et al. \cite{Cuperman2009-fp} showed that participants with high agreeableness tend to rate rapport toward their conversational partners highly and to be rated highly by their conversational partners. Wood et al. \cite{Wood2010-sa} also reported that participants with high agreeableness tend to rate their friend positively. Furthermore, participants' extraversion and agreeableness are positively correlated with their ratings of friendship satisfaction \cite{Harris2016-xv}. Overall, the Big Five dimensions most closely related to interpersonal perception are extraversion and agreeableness, but other dimensions can influence it as well. Inspired by these studies, we hypothesize that personality trait features promote estimation models to capture people's rating tendencies. We introduce SRM to confirm this hypothesis. Our study is the first attempt to use SRM to analyze estimation models.

\begin{figure}[t]
    \centering
    \includegraphics[width=1.0\linewidth]{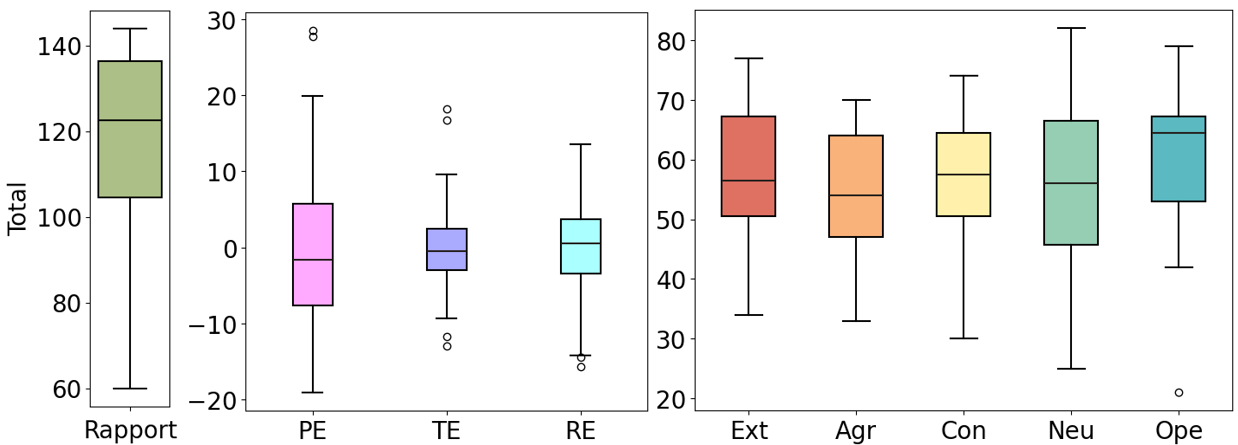}
    \caption{Box plots of rapport scores and the relationship effects ($N$ = 96), the perceiver and the target effects, and Big Five dimensions ($N$ = 32). PE = Perceiver Effect, TE = Target Effect, RE = Relationship Effect, Ext = Extraversion, Agr = Agreeableness, Con = Conscientiousness, Neu = Neuroticism, Ope = Openness.}
    \label{fig:response}
\end{figure}

\section{DATA}
\label{Data}
To achieve our research objectives, we collected online dyadic interactions between friends. In Section \ref{Dataset}, we present an overview of our dataset, and in Section \ref{Data Analysis}, we analyze the relationship between the Big Five and the three effects of rapport ratings. All recordings were reviewed and approved by our institution's research ethics committee.

\subsection{Dataset}
\label{Dataset}
We recruited eight groups consisting of four participants each who were friends with each other through a recruitment agency. There were a total of 32 Japanese participants (16 men, 16 women; 26 in their 20s, 6 in their 30s), and each participant was paired with other participants in the same group (48 dyads). All participants in the same dyads were the same gender. Each dyad conducted three online interactions based on different conversation topics. Each interaction lasted 20 minutes. We used the first interactions for our analysis and experiments because the first are the least restrictive and the most natural interactions. In the first interaction, the dyad introduces themselves (e.g., their favorite foods and artists). For more information about the recording settings and topics, see Section I\hspace{-1.2pt}I\hspace{-1.2pt}I of \cite{Hayashi2023-bp}, as we strictly followed their procedure for collecting initial interactions.

\begin{figure}[t]
    \centering
    \includegraphics[width=1.0\linewidth]{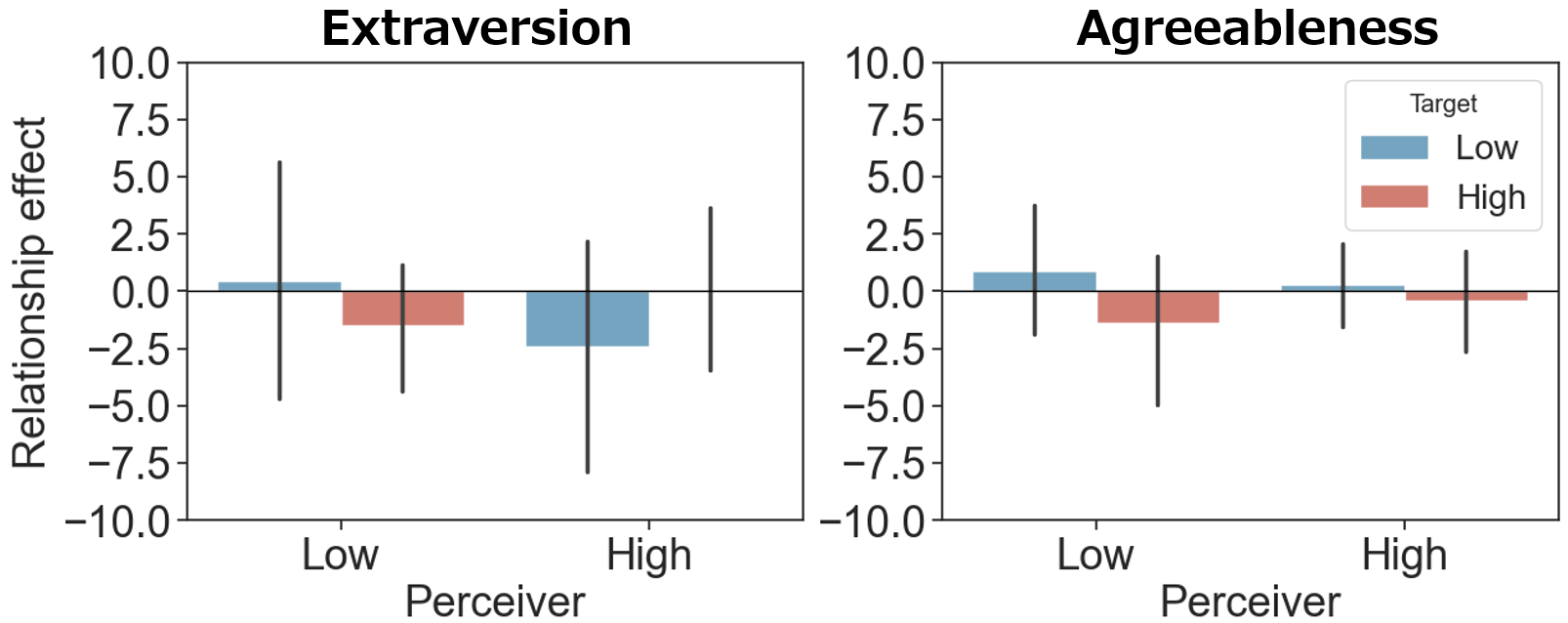}
    \caption{Relationship effects according to combinations of Big Five dimensions. Error bars represent standard deviation.}
    \label{fig:boxplot}
\end{figure}

The dataset includes participants’ rapport ratings for their conversational partners, which was reported after each interaction. We utilized a questionnaire developed by Bernieri et al. \cite{Bernieri1996-it} to measure self-reported rapport. A previous study translated the questionnaire from English into Japanese, and its internal consistency among items was sufficient ($\alpha=0.92$) \cite{Kimura2005-dw}. The questionnaire comprises 18 items, each of which is rated on an 8-point Likert scale (1 = strongly disagree, 8 = strongly agree). We define the rapport score as the sum of the responses after reversing the values of the negative questions. In addition, at the end of all recordings, participants completed the Japanese Big Five questionnaire \cite{Wada1996-pk}, which consists of 60 items on a 7-point Likert scale. After reversing the values of the negative questions, we summed the responses along each Big Five dimension; Fig. \ref{fig:boxplot} shows box plots of the rapport score, the three effects, and the Big Five dimensions.

\begin{table}[t]
  \caption{Associations of perceiver/target effects with Big Five dimensions.}
  \label{tab:Per_Tar_Effect}
  \begin{center}
  \begin{tabular}{c|ccccc}
    \toprule
    &Ext&Agr&Con&Neu&Ope\\
    \midrule
    Perceiver eff.&.03&.15&--.06&.25&.08\\
    Target eff.&.21&.10&--.21&.15&.21\\
  \bottomrule
\end{tabular} \\
\end{center}
\end{table}

\subsection{Data Analysis}
\label{Data Analysis}
\subsubsection{Perceiver effect and target effect}
We calculated the Pearson's product-moment correlation coefficient to estimate the association between the perceiver/target effects and Big Five dimensions (see Table \ref{tab:Per_Tar_Effect}). In a previous study conducting large-scale research on initial interactions (87 dyads), extraversion and agreeableness were significantly associated with perceiver and target effects of rapport perception \cite{Cuperman2009-fp}. Although we found no significant correlations in our data (significance level $\alpha$ = .05), this result may be due to the insufficient sample size (48 dyads). A small positive correlation was observed between the perceiver effects and neuroticism (.25). Extraversion and openness had small positive correlations with the target effects (.21), and conscientiousness had a small negative correlation with the target effects (--.21). Our different results here compared to previous studies \cite{Cuperman2009-fp} may be due to differences in cultural backgrounds or relationships among participants, but we cannot say this for certain. Overall, the findings suggest that the perceiver's and target's Big Five features (BFFs) lead models to capture the perceiver and target effects, respectively. 

\subsubsection{Relationship effect}
We present bar plots in Fig. \ref{fig:boxplot} to visualize the relationship effects according to dyads that have specific combinations of Big Five dimensions. For each Big Five dimension, we categorized the perceiver and target into two types (High or Low) using a threshold. The threshold is defined as $M \pm SD \times 0.5$, where $M$ is the mean total responses corresponding to the Big Five dimension and $SD$ is the standard deviation. We focus here on extraversion and agreeableness, as combinations of type in these dimensions lead to unique high or low rapport \cite{Cuperman2009-fp}. In our dataset, dyads with different extraversion types tended to yield negative relationship effects. Furthermore, dyads between a disagreeable perceiver and an agreeable target yielded negative relationship effects. These results suggest that the combination of the perceiver's and target's BFFs leads models to accurately capture the relationship effects.

\begin{figure*}[t]
    \centering
    \includegraphics[width=0.8\linewidth]{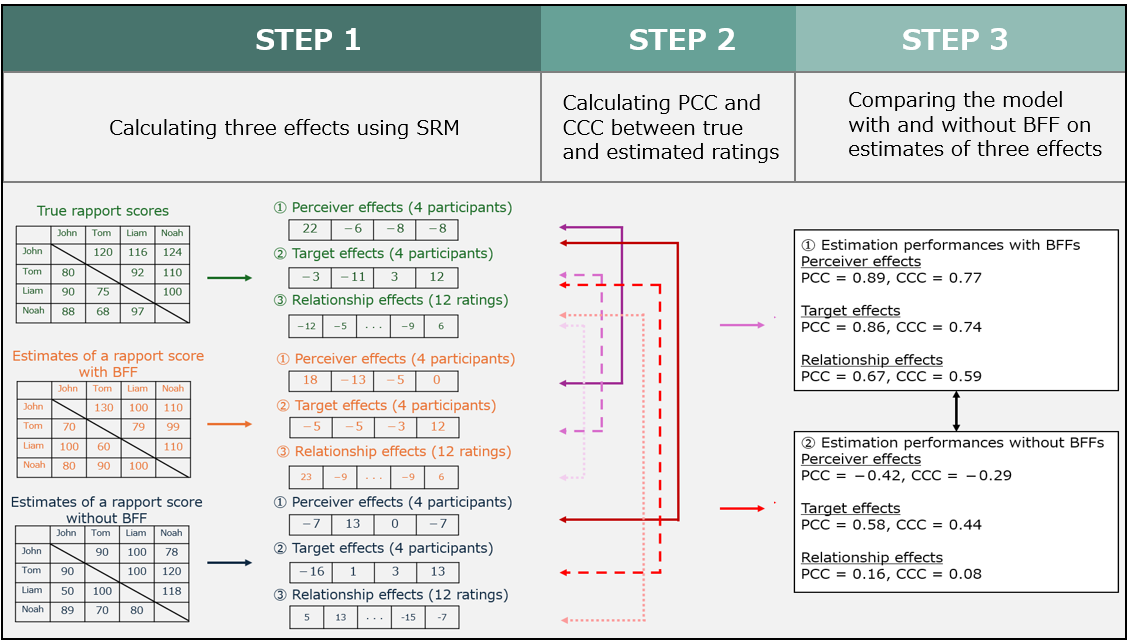}
    \caption{Our Proposed Analytical approach based on SRM}
    \label{fig:AnalyticalApproach}
\end{figure*}

\section{METHOD}
\label{Method}
\label{Method}
\subsection{Features Extraction}
\subsubsection{Audio features}
We used OpenSMILE \cite{Eyben2010-zx} to extract audio features from each utterance. Audio features were based on eGeMAPS \cite{Eyben2016-ax}, which is recognized as the default setting in speech emotion recognition. Audio features were 88 dimensions, including pitch and volume. We standardized them across the utterance pool of each participant.

\subsubsection{Facial expression features}
We utilized OpenFace \cite{Baltrusaitis2018-yj} to extract the intensity of 17 action units (AUs) from each frame. Facial expression features included 14 statistics calculated for each AU; thus, they totaled 238 ($17\times14$) dimensions. The 14 statistics were the mean, median, standard deviation, skewness, kurtosis, maximum and minimum, mean of the first and second differences, range, slope, intercept of the linear approximation, and the 25th and 75th percentile values. We standardized the facial expression features across the utterance pool of each participant.

\subsubsection{Big Five features (BFFs)}
We utilized responses on the Big Five questionnaire as BFFs. The Big Five questionnaire quantifies the Big Five dimensions of Extraversion, Agreeableness, Conscientiousness, Neuroticism, and Openness. The BFFs had 60 dimensions, which we normalized across the participant pool using min-max normalization. 

\subsection{Estimation Model}
\subsubsection{Problem definition}
The problem was to estimate the perceiver's rapport scores as regression. Take the mapping function $f:\bm{X} \rightarrow y$. The $y$ is the rapport score reported by the perceiver. The $\bm{X}$ consists of the sequence of perceiver's utterance-level features and BFFs: $\bm{X} = \{\bm{U_1},$ $\cdots, \bm{U_T}, \bm{B} \}$, where $T$ is the total number of perceiver's utterances. In a unimodal setting, each utterance-level features comprises a single modality with $\bm{U_i} \in \mathbb{R}^{1 \times D_U}$, where $D_U$ is the dimension of unimodal features. In a bimodal setting, $D_U$ represents the total dimension of two unimodal features. The $\bm{B} \in \mathbb{R}^{1 \times D_B}$ is BFFs, where $D_B$ represents the dimension of BFFs.

\subsubsection{Model architecture and training}
We developed a mapping function $f$ inspired by Poria et al.\cite{Poria2017-nf}. Our mapping function is composed of unidirectional long short-term memory networks (LSTM) and fully connected neural networks (FCNN). The perceiver's sequence of utterances $\{\bm{U_1},$ $\cdots, \bm{U_T} \}$ is input to the LSTM, and the output vector corresponding to the last utterance $\bm{h}_T$ is extracted. We then concatenate $\bm{h}_T$ and BFFs $\bm{B}$, and map this vector to the estimate of rapport score $\hat{y}$,
\begin{equation}
\bm{h}_T = \mathrm{LSTM}(\bm{U_1}, \cdots, \bm{U_T}),
\end{equation}
\begin{equation}
\hat{y} = \mathrm{FCNN}(\bm{h}_T \oplus \bm{B}).
\end{equation}
The model is trained by minimizing the mean square error.

\subsection{Analytical Approach using Social Relations Model (SRM)}
Related to RO-2, we explain how to investigate the role of BFFs in performance improvement using SRM. As shown in Eq. (1), the rapport score consist of the perceiver, target, relationship effect, and group average. Thus, when the estimation performance of rapport score, the model becomes able to capture any or all of these effects more accurately. Our approach is designed to demystify the performance improvement of the rapport score based on the variation in the estimation performance of each effect. This approach consists of three steps (see Fig. \ref{fig:AnalyticalApproach}).

In the first step, we decompose true (human) rapport scores and estimates of rapport scores into perceiver, target, and relationship effects using SRM. We compute the three effects from the true rapport score according to Eqs. (2)--(4). We define these effects as true effects. We also compute the three effects of the estimated rapport score. We define these effects as the estimates of effects.

In the second step, we calculate the Pearson's correlation coefficient (PCC) and concordance correlation coefficient (CCC) between the true effects and the estimates of effects.

In the third step, we compare the PCC and CCC for each effect between models with and without BFFs. By comparing the two models, it becomes quantitatively clear how useful the BFFs are in accurately capturing each effect.

\begin{table*}[t]
\caption{Estimation performances of rapport scores and perceiver, target, and relationship effects.}
\label{tab:Main_result}
\centering
\scalebox{1.0}{
\begin{tabular}{c|cc|cc|cc|cc}
\hline
 & \multicolumn{2}{c|}{Rapport Score} &  \multicolumn{2}{c|}{Perceiver Eff.} & \multicolumn{2}{c|}{Target Eff.} & \multicolumn{2}{c}{Relationship Eff.} \\
Modality & PCC $\uparrow$ & CCC $\uparrow$ &  PCC $\uparrow$ & CCC $\uparrow$ & PCC $\uparrow$ & CCC $\uparrow$ & PCC $\uparrow$ & CCC $\uparrow$ \\ \hline  \hline
A & .05 \tiny{$\pm.04$} & .03 \tiny{$\pm.03$} & --.29 \tiny{$\pm.07$} & --.26 \tiny{$\pm.05$} & .17 \tiny{$\pm.09$} & .09 \tiny{$\pm.08$} & .19 \tiny{$\pm.08$} & .17 \tiny{$\pm.07$} \\
+BFF [P] & $.10^{\ast}$ \tiny{$\pm.06$} & .04 \tiny{$\pm.04$} & $-.04^{\ast}$ \tiny{$\pm.10$} & $\bf-.06^{\ast}$ \tiny{$\pm.08$} & .12 \tiny{$\pm.08$} & .04 \tiny{$\pm.08$} & $.25^{\ast}$ \tiny{$\pm.07$} & $.24^{\ast}$ \tiny{$\pm.07$} \\
+BFF [T] & .07 \tiny{$\pm.04$} & $.05^{\ast}$ \tiny{$\pm.02$} & --.28 \tiny{$\pm.09$} & --.25 \tiny{$\pm.06$} & .21 \tiny{$\pm.11$} & $\bf.12^{\ast}$ \tiny{$\pm.08$} & .23 \tiny{$\pm.06$} & $.21^{\ast}$ \tiny{$\pm.05$} \\
+BFF [P+T] & $\bf{.12}^{\ast}$ \tiny{$\pm.04$} & $\bf.06^{\ast}$ \tiny{$\pm.03$} & $\bf{-.02}^{\ast}$ \tiny{$\pm.07$} & $-.07^{\ast}$ \tiny{$\pm.06$} & $\underline{\bf.23}^{\ast}$ \tiny{$\pm.08$} & .11 \tiny{$\pm.07$} & $\bf.29^{\ast}$ \tiny{$\pm.08$} & $\bf{.27}^{\ast}$ \tiny{$\pm.07$} \\ \hline
F & .31 \tiny{$\pm.05$} & .18 \tiny{$\pm.03$} & .20 \tiny{$\pm.12$} & .18 \tiny{$\pm.10$} & -.01 \tiny{$\pm.07$} & .02 \tiny{$\pm.05$} & .41 \tiny{$\pm.05$} & .38 \tiny{$\pm.05$} \\
+BFF [P] & .32 \tiny{$\pm.05$} & .18 \tiny{$\pm.03$} & $.30^{\ast}$ \tiny{$\pm.11$} & $.25^{\ast}$ \tiny{$\pm.08$} & $.05^{\ast}$ \tiny{$\pm.08$} & $.09^{\ast}$ \tiny{$\pm.05$} & $\underline{\bf.45}^{\ast}$ \tiny{$\pm.07$} & \underline{\bf.40} \tiny{$\pm.07$} \\
+BFF [T] & $.35^{\ast}$ \tiny{$\pm.04$} & $.20^{\ast}$ \tiny{$\pm.02$} & $.27^{\ast}$ \tiny{$\pm.10$} & $.22^{\ast}$ \tiny{$\pm.08$} & $.15^{\ast}$ \tiny{$\pm.10$} & $.14^{\ast}$ \tiny{$\pm.07$} & .41 \tiny{$\pm.06$} & .38 \tiny{$\pm.06$} \\
+BFF [P+T] & $\underline{\bf.38}^{\ast}$ \tiny{$\pm.04$} & $\underline{\bf.21}^{\ast}$ \tiny{$\pm.03$} & $\underline{\bf.36}^{\ast}$ \tiny{$\pm.09$} & $\underline{\bf.28}^{\ast}$ \tiny{$\pm.07$} & ${\bf.17}^{\ast}$ \tiny{$\pm.08$} & $\underline{\bf.16}^{\ast}$ \tiny{$\pm.06$} & .43 \tiny{$\pm.06$} & {.39} \tiny{$\pm.06$} \\ \hline
A+F & .24 \tiny{$\pm.05$} & .12 \tiny{$\pm.03$} & .06 \tiny{$\pm.11$} & .04 \tiny{$\pm.10$} & .02 \tiny{$\pm.11$} & .06 \tiny{$\pm.08$} & \bf{.40} \tiny{$\pm.06$} & \bf.36 \tiny{$\pm.06$} \\
+BFF [P] & .21 \tiny{$\pm.05$} & .10 \tiny{$\pm.03$} & $.11^{\ast}$ \tiny{$\pm.12$} & $.09^{\ast}$ \tiny{$\pm.09$} & .02 \tiny{$\pm.09$} & .08 \tiny{$\pm.07$} & .36 \tiny{$\pm.07$} & .32 \tiny{$\pm.05$} \\
+BFF [T] & $\bf.26^{\ast}$ \tiny{$\pm.03$} & \bf.13 \tiny{$\pm.02$} & .07 \tiny{$\pm.07$} & .05 \tiny{$\pm.06$} & $\bf.18^{\ast}$ \tiny{$\pm.10$} & $\bf.15^{\ast}$ \tiny{$\pm.06$} & .34 \tiny{$\pm.07$} & .31 \tiny{$\pm.06$} \\
+BFF [P+T] & .25 \tiny{$\pm.06$} & {.13} \tiny{$\pm.04$} & $\bf{.15}^{\ast}$ \tiny{$\pm.10$} & $\bf{.10}^{\ast}$ \tiny{$\pm.09$} & $.14^{\ast}$ \tiny{$\pm.13$} & $.13^{\ast}$ \tiny{$\pm.09$} & .37 \tiny{$\pm.07$} & .32 \tiny{$\pm.06$} \\ \hline
BFF [P] & .01 \tiny{$\pm.09$} & .00 \tiny{$\pm.04$} & .07 \tiny{$\pm.11$} & .02 \tiny{$\pm.09$} & --- & ---  & --- & --- \\
BFF [T] & .06 \tiny{$\pm.04$} & .03 \tiny{$\pm.02$} & --- & --- & .17 \tiny{$\pm.07$} & .12 \tiny{$\pm.06$} & --- & --- \\
BFF [P+T] & .09 \tiny{$\pm.06$} & .04 \tiny{$\pm.04$} & .12 \tiny{$\pm.08$} & .06 \tiny{$\pm.06$} & .17 \tiny{$\pm.11$} & .10 \tiny{$\pm.09$} & .14 \tiny{$\pm.16$} & .10 \tiny{$\pm.08$} \\ \hline
Random & .00 \tiny{$\pm.12$} & {-.00} \tiny{$\pm.07$} & {-.02} \tiny{$\pm.19$} & {-.02} \tiny{$\pm.16$} & {-.03} \tiny{$\pm.18$} & -.01 \tiny{$\pm.15$} & .03 \tiny{$\pm.14$} & .01 \tiny{$\pm.09$} \\ \hline
\end{tabular}
}\\
A = Audio features; F = Facial expression features; BFF = Big Five features; P = Perceiver; T = Target. \\ Bold and underlined values represent the best performances within each modality and across modalities, respectively. \\ The asterisk denotes that the performance is significantly better than that of the model without BFF for each modality.
\end{table*}

\section{EXPERIMENTAL SETTINGS}
\label{experimental_settings}
\subsection{Evaluation Methods}
We evaluate estimation performances using an eight-fold cross-validation, where each fold corresponds to one group consisting of four participants. The cross-validation ensured that the same participant was not duplicated across the training and test sets. In our data, 96 samples were created from 48 dyads since the rapport rating is bidirectional. 

The Pearson's correlation coefficient (PCC) and concordance correlation coefficient (CCC) \cite{Lin1989ACC} were calculated based on true rapport scores and estimated rapport scores. PCC ($\rho$) is defined as

\begin{equation}
\rho =  \frac{\sigma_{y\hat{y}}}{\sigma_{y}\sigma_{\hat{y}}},
\end{equation}
where $\sigma_{y}$ and $\sigma_{\hat{y}}$ are standard deviations of true and estimated rapport scores, respectively. $\sigma_{y\hat{y}}$ is their covariance. CCC ($\rho_c$) is defined as
\begin{equation}
\rho_c =  C\rho,
\end{equation}
where C is a bias correction factor:
\begin{equation}
C =  \frac{2\sigma_{y}\sigma_{\hat{y}}}{\sigma^{2}_{y}+\sigma^{2}_{\hat{y}}+(\mu_y - \mu_{\hat{y}})^2}.
\end{equation}
The $C$ is between 0 and 1; the maximum value is realized when two variables have an identical mean and standard deviation. Thus, the CCC is a correlation coefficient that takes into account the similarity of distribution between two variables. Each experiment was performed 30 times based on different random seed values, and we reported the average performance and standard deviation to reduce the influence of the initial model parameters.

\subsection{Model Settings}
For training on the estimation models, we set the mini-batch size to 32 and the number of epochs to 50. Dropout was applied to the hidden layer of FCNN with a drop rate of 30\%. The learning rate was 1.5e-4 for the models using audio features, 1.0e-4 for the models using facial expression features and bimodal features, and 1.0e-3 for the models using only BFFs to ensure that all models were sufficiently fitted to the training data with 50 epochs. All models were implemented in Pytorch 2.0.1, and all experiments were conducted on NVIDIA GeForce RTX 3090.

\section{RESULTS}
\label{RESULT}
Table \ref{tab:Main_result} shows the estimation performance of rapport scores and the perceiver, target, and relationship effects. Bold and underlined values represent the best performances within each modality and across modalities, respectively. The standard deviation of the rapport score and relationship effect represents the amount of variation per rapport score; the standard deviation of the perceiver and target effect represents the amount of variation per person. As a naive baseline, we used a model that predicts values randomly within the range of true rapport scores. To test if the performances of each estimation model are significantly better than those of the baseline, we conducted a Mann–Whitney U test ($N$=30, $\alpha$=.05, one-sided test), where $N$ corresponds to the number of random seed values. We also conducted the same statistical test to determine if the performance with Big Five features (BFFs) was significantly better than that without BFFs for each modality; the asterisk denotes this significant differences. In this section, we first report the performance of rapport scores. Then, we review the performance of the three effects. We mainly focus on unimodal models, as these models are the minimal units of bimodal models.

\subsection{Estimation Performance of Rapport Scores}
Rows 2 and 3 in Table~\ref{tab:Main_result} list the estimation performances of the rapport scores. Overall, models with facial expressions achieved high performances. Conversely, those with audio were low. In both modalities, the statistical test showed that the performances of all models were significantly better than that of the random baseline. Therefore, models were able to sufficiently predict rapport scores.

The results also indicate that adding BFFs improves performances. In facial expressions, the model (F + BFF[P+T]) achieved the highest performances across modalities with PCC (.38) and CCC (.21). Furthermore, both of the BFF[P] alone and the BFF[T] alone yielded performance gains. In audio, BFFs also led to better performances. In short, almost all models with BFFs achieved significantly better performances than models without BFFs in both modalities.

\subsection{Implicit Estimation Performance of Perceiver, Target, and Relationship Effects}
\label{RESULT:effect}
Next, we examine how BFFs contribute to capturing the three effects. Rows 4 to 9 in Table~\ref{tab:Main_result} list the implicit estimation performances of the three effects. The term \textit{implicit} here denotes that models do not estimate the three effects directly; rather, the estimates are calculated using a post--hoc approach. In both modalities, the results of the statistical test showed that models achieved significantly better performances than the random baseline except in these situations: PCC and CCC for perceiver effects by all models using audio; PCC for target effects by the model (F alone).

Regarding the \textit{perceiver effect}, experimental results indicated that adding the perceiver's BFFs improves the performances of perceiver effects. The overall performance of models using facial expression was better than that of the other models. Specifically, model (F + BFF[P+T]) achieved the best performances across all modalities (PCC: .36, CCC: 0.28). Models (A + BFF[P+T]) and (A + BFF[P]) also accomplished the best PCC and CCC within the audio, respectively. Although BFF[T] improved performances in both modalities, the performance gain by BFF[P] was substantially better than that by BFF[T].

Regarding the \textit{target effect}, the results indicated that adding the target's BFFs improves the performances of target effects. The overall performance of the audio model was better than that of the other models. The model (A+BFF[P+T]) achieved the highest PCC across modalities (.23), and Model (F+BFF[P+T]) achieved the highest CCC across modalities (.16). In audio, BFF[T] improved performances, although BFF[P] did not. In facial expressions, both BFF[P] and BFF[T] yielded performance gains. However, the performance gain by BFF[T] (PCC: +.16, CCC: +.12) was substantially better than that by BFF[P] (PCC: +.06, CCC: +.07).

Regarding the \textit{relationship effect}, the results indicated that adding BFFs improves the performances of relationship effects. The overall performance of models using facial expressions was higher than that of the other models. Model (A+BFF[P]) achieved the highest performances across all modalities (PCC: .45, CCC: .40). BFF[P+T] yielded the best performances within audio; however, it did not yield the best performance within facial expression. 

\begin{figure*}[t]
    \centering
    \includegraphics[width=0.90\linewidth]{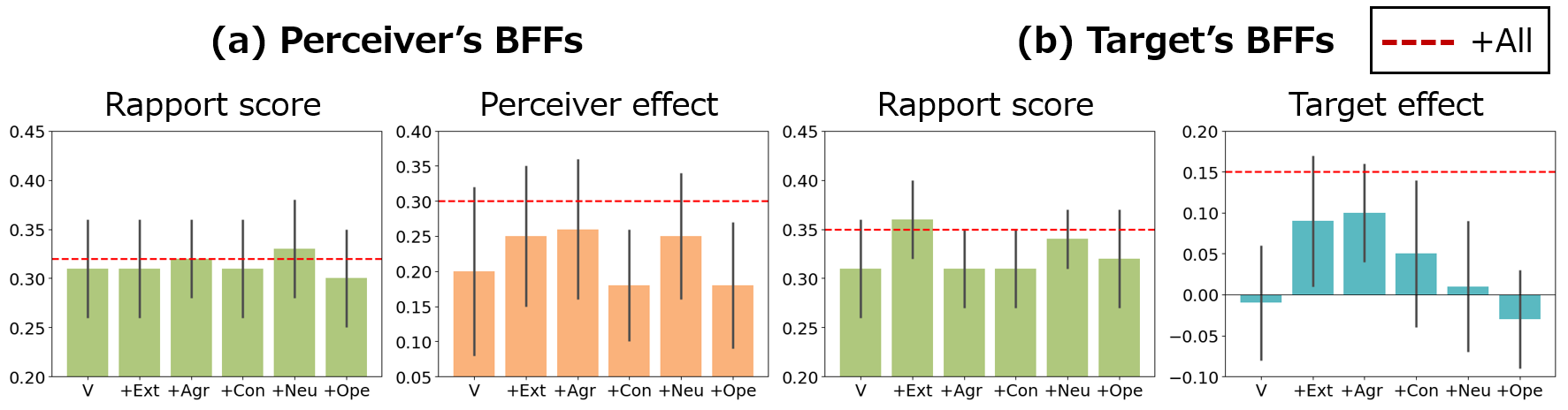}
    \caption{Relationship between PCC and Big Five dimensions (BFDs). In (a), we investigate the effectiveness of the perceiver's BFDs for improving the PCC of rapport score and perceiver effects. In (b), we explore the effectiveness of the target's BFDs for improving the PCC of rapport score and target effects. The red dotted line represents the PCC of models with all BFDs.}
    \label{fig:Bf_dimenstions}
\end{figure*}

\section{DISCUSSION}
\label{Discussion}
We first discuss whether Big Five features (BFFs) can improve the estimation performance of rapport in interactions between friends (\textbf{RO-1}). We then examine the role of BFFs in performance improvement (\textbf{RO-2}). Next, we investigate which Big Five dimensions (BFDs) are effective in rapport estimation. Finally, we touch on the limitations of our study and the versatility of our proposed analytical approach.

\subsection{Big Five features improve rapport estimation performance between friends (RO-1).}
The experimental results showed that facial expressions contain rich information for rapport estimation. This finding is in line with a previous study \cite{Muller2018-ax} revealing that facial expression features yielded better estimation performances of rapport than other nonverbal features (e.g., audio features). The results also highlighted the importance of the BFFs for rapport estimation not only in initial interactions \cite{Cerekovic2017-tq, Muller2018-ax} but also in interactions between friends. Specifically, combining the perceiver's and the target's BFFs yields the best performance. This result makes sense because rapport is built among two participants, and their Big Five personality traits influence rapport building \cite{Cuperman2009-fp}. 

The implication of these findings is that collecting the target's Big Five personality traits is useful for estimation models with facial expression to substantially improve their performance. However, practitioners do not always have access to it in practical situations. Therefore, further studies is required to determined actual personality trait can be substituted with personality trait estimated by estimation models (e.g., \cite{Mawalim2023-sq}) for the performance improvement. 

\subsection{BFFs promote models to capture three effects (RO-2).}
We verify three hypotheses: 1) the perceiver's BFFs lead models to capture perceiver effects, 2) the target's BFFs lead models to capture target effects, and 3) combinations of the two lead models to capture relationship effects. 

First, the experimental results showed that BFF[P] lead models to capture \textit{perceiver effects}, which substantiates our first hypothesis. The performance gain by BFF[P] was better than that by BFF[T]. The results can be explained by evidence that the perceiver’s Big Five personality traits are more strongly associated with the perceiver’s tendency to rate other people \cite{Cuperman2009-fp, Funder1993-xi}. Furthermore, adding BFF[T] also yielded performance gains. Although it is not clear how the target’s BFFs contributed to the performance gain, one possibility is that the perceiver's nonverbal behavior toward a conversation partner with specific Big Five personality traits contains rich information about perceiver effects. The performance gain by adding BFFs[P+T] (e.g., PCC: .16, CCC: .10 in facial expression) is greater than that of models using BFFs[P+T] alone. Thus, nonverbal features and BFFs may have a synergistic effect for capturing perceiver effects.

Second, we confirmed that BFF[T] leads models to capture \textit{target effects}, which supports our second hypothesis. The performance gain by BFF[T] was better than that by BFF[P]. This result is in line with findings showing that the target’s Big Five more strongly influences the target’s tendency to be rated by other people \cite{Cuperman2009-fp, Funder1993-xi}. Furthermore, BFF[P] also yielded better performances in facial expression. Although we cannot say for certain how the perceiver’s BFFs contributed to the performance gain, one possible explanation is that the facial expression by the perceiver with specific Big Five personality traits may be an indicator of target effects.

Finally, BFF[P+T] leads models to capture \textit{relationship effects} in audio, which partially confirms our third hypothesis. In audio, BFF[P+T] led to the best PCC and CCC. However, in facial expression, BFF[P+T] did not yielded the best performances. The result imply that the perceiver's audio features and both BFFs are complementary relationship for capture the unique rapport of dyads.

Next, we focus on relationship between modality and the estimation performances of three effects. On the whole, facial expression lead models better estimation performances of the perceiver and the relationship effects than audio; in contrast, audio bring models better estimation performances of the target effects. One possible explanation for the results is that specific facial expression (e.g., smile) is associated with the tendency to feel rapport for other people and unique rapport. On the other hand, specific prosody may be useful cues of the tendency to be felt rapport by other people.

The implication of these findings is that BFFs yield better performances of rapport because the perceiver's and the target's BFFs help models to capture the corresponding effects, respectively. Furthermore, the findings clarify which types of BFF are effective for obtaining the desired auxiliary information for rapport. If practitioners need to detect people who are exceptionally misanthropic (i.e., low perceiver effect) or who are undesirable (i.e., low target effect) \cite{Joel2017-ne}, the utilization of the perceiver's BFFs or the target's BFFs, respectively, is a suitable choice.

\subsection{Analysis of Effective Big Five Dimensions}
Fig. \ref{fig:Bf_dimenstions} shows the relationship between the PCC and BFDs. This analysis focuses only on facial expression features due to their high estimation performances. This analysis also focuses on PCC, as PCC and CCC have a similar ordinal relationship according to the type of features. We investigate the effectiveness of the perceiver's and the target's BFDs for improving the PCC of rapport scores and the corresponding effects. The red dotted line represents the PCC of models with all BFDs. The error bar denotes the standard deviations between 30 random seed values.

In (a), we examine the effectiveness of the perceiver's BFDs for enhancing the PCC. The perceiver's agreeableness and neuroticism improved the PCC of rapport scores; the effective perceiver's BFDs for improving perceiver effects were extraversion, agreeableness, and neuroticism. Agreeableness and neuroticism corresponded to BFDs that highly correlated with perceiver effects (see Table \ref{tab:Per_Tar_Effect}).

In (b), we explore the effectiveness of the target's BFDs for enhancing the PCC. The target's extraversion, agreeableness, and neuroticism improved the PCC of rapport scores. Furthermore, the effective target's BFDs for estimating target effects were extraversion, agreeableness, and conscientiousness. Extraversion and conscientiousness corresponded to BFDs that highly correlated with target effects (see Table \ref{tab:Per_Tar_Effect}). However, the target's openness yielded performance loss despite its high correlations with target effects.

The BFDs which strongly effects to interpersonal perception are extraversion and agreeableness \cite{Cuperman2009-fp, Harris2016-xv}. The results revealed that these BFDs are also important cues for implicitly effect estimation. Furthermore, specific BFDs did not reach the estimation performances with all BFDs (red dotted line) on perceiver and target effects. These findings indicate that implicit capturing of perceiver and target effects is based not on specific BFDs but rather on multiple BFDs.

\subsection{Limitations}
As with most studies related to interpersonal perception, our findings have limitations in their generality. Specifically, our study imposed limitations on the generality of the findings with respect to the participants’ restricted range of age and cultural backgrounds. In addition, we relied on limited conversation settings, such as online conversations. Therefore, further research is needed to determine whether our findings can be generalized to various participants and conversation settings. Despite these limitations, our study represents an important first step to understanding the role of personality traits in interpersonal perception.

\subsection{Versatility of our Analytical Approach}
Beyond the limit of rapport and BFFs, our analysis approach based on the social relations model (SRM) is helpful in analyzing the extent to which model improvements and data extensions contribute to capturing different aspects of interpersonal perception. Our approach can be applied to interpersonal perception other than rapport (e.g., romantic desire) and features other than BFFs (e.g., attachment style). In summary, our approach is an effective tool for understanding personality-aware estimation models of interpersonal perception and obtaining insight to improve them.

\section{CONCLUSION}
\label{Conclusion}
We (1) investigated whether Big Five features (BFFs) can improve the estimation performance of rapport in interactions between friends and (2) examined the role of BFFs in improving the estimation performance of rapport. Our experimental results showed that BFFs substantially improve the estimation performance of rapport in interactions between friends. Furthermore, we demonstrated that BFFs yield better estimation performances of rapport, as the perceiver's and target's BFFs lead models to accurately capture the corresponding effects. Our study is the first step toward understanding why personality-aware estimation models achieve high estimation performances of interpersonal perception.



\bibliographystyle{ACM-Reference-Format}
\bibliography{main}


\begin{thebibliography}{32}


\ifx \showCODEN    \undefined \def \showCODEN     #1{\unskip}     \fi
\ifx \showDOI      \undefined \def \showDOI       #1{#1}\fi
\ifx \showISBNx    \undefined \def \showISBNx     #1{\unskip}     \fi
\ifx \showISBNxiii \undefined \def \showISBNxiii  #1{\unskip}     \fi
\ifx \showISSN     \undefined \def \showISSN      #1{\unskip}     \fi
\ifx \showLCCN     \undefined \def \showLCCN      #1{\unskip}     \fi
\ifx \shownote     \undefined \def \shownote      #1{#1}          \fi
\ifx \showarticletitle \undefined \def \showarticletitle #1{#1}   \fi
\ifx \showURL      \undefined \def \showURL       {\relax}        \fi
\providecommand\bibfield[2]{#2}
\providecommand\bibinfo[2]{#2}
\providecommand\natexlab[1]{#1}
\providecommand\showeprint[2][]{arXiv:#2}

\bibitem[Baltrusaitis et~al\mbox{.}(2018)]%
        {Baltrusaitis2018-yj}
\bibfield{author}{\bibinfo{person}{Tadas Baltrusaitis}, \bibinfo{person}{Amir Zadeh}, \bibinfo{person}{Yao~Chong Lim}, {and} \bibinfo{person}{Louis-Philippe Morency}.} \bibinfo{year}{2018}\natexlab{}.
\newblock \showarticletitle{{OpenFace} 2.0: Facial Behavior Analysis Toolkit}. In \bibinfo{booktitle}{\emph{2018 13th {IEEE} International Conference on Automatic Face \& Gesture Recognition ({FG} 2018)}}. \bibinfo{publisher}{IEEE}, \bibinfo{pages}{59--66}.
\newblock


\bibitem[Bernieri et~al\mbox{.}(1996)]%
        {Bernieri1996-it}
\bibfield{author}{\bibinfo{person}{Frank~J Bernieri}, \bibinfo{person}{John~S Gillis}, \bibinfo{person}{Janet~M Davis}, {and} \bibinfo{person}{Jon~E Grahe}.} \bibinfo{year}{1996}\natexlab{}.
\newblock \showarticletitle{Dyad rapport and the accuracy of its judgment across situations: A lens model analysis}.
\newblock \bibinfo{journal}{\emph{J. Pers. Soc. Psychol.}} \bibinfo{volume}{71}, \bibinfo{number}{1} (\bibinfo{date}{July} \bibinfo{year}{1996}), \bibinfo{pages}{110--129}.
\newblock


\bibitem[Cerekovic et~al\mbox{.}(2017)]%
        {Cerekovic2017-tq}
\bibfield{author}{\bibinfo{person}{Aleksandra Cerekovic}, \bibinfo{person}{Oya Aran}, {and} \bibinfo{person}{Daniel Gatica-Perez}.} \bibinfo{year}{2017}\natexlab{}.
\newblock \showarticletitle{Rapport with Virtual Agents: What Do Human Social Cues and Personality Explain?}
\newblock \bibinfo{journal}{\emph{IEEE Transactions on Affective Computing}} \bibinfo{volume}{8}, \bibinfo{number}{3} (\bibinfo{year}{2017}), \bibinfo{pages}{382--395}.
\newblock


\bibitem[Cuperman and Ickes(2009)]%
        {Cuperman2009-fp}
\bibfield{author}{\bibinfo{person}{Ronen Cuperman} {and} \bibinfo{person}{William Ickes}.} \bibinfo{year}{2009}\natexlab{}.
\newblock \showarticletitle{Big Five predictors of behavior and perceptions in initial dyadic interactions: personality similarity helps extraverts and introverts, but hurts ``disagreeables''}.
\newblock \bibinfo{journal}{\emph{J. Pers. Soc. Psychol.}} \bibinfo{volume}{97}, \bibinfo{number}{4} (\bibinfo{date}{Oct.} \bibinfo{year}{2009}), \bibinfo{pages}{667--684}.
\newblock


\bibitem[Eyben et~al\mbox{.}(2016)]%
        {Eyben2016-ax}
\bibfield{author}{\bibinfo{person}{Florian Eyben}, \bibinfo{person}{Klaus~R Scherer}, \bibinfo{person}{Bj{\"o}rn~W Schuller}, \bibinfo{person}{Johan Sundberg}, \bibinfo{person}{Elisabeth Andr{\'e}}, \bibinfo{person}{Carlos Busso}, \bibinfo{person}{Laurence~Y Devillers}, \bibinfo{person}{Julien Epps}, \bibinfo{person}{Petri Laukka}, \bibinfo{person}{Shrikanth~S Narayanan}, {and} \bibinfo{person}{Khiet~P Truong}.} \bibinfo{year}{2016}\natexlab{}.
\newblock \showarticletitle{The Geneva Minimalistic Acoustic Parameter Set ({GeMAPS}) for Voice Research and Affective Computing}.
\newblock \bibinfo{journal}{\emph{IEEE Transactions on Affective Computing}} \bibinfo{volume}{7}, \bibinfo{number}{2} (\bibinfo{year}{2016}), \bibinfo{pages}{190--202}.
\newblock


\bibitem[Eyben et~al\mbox{.}(2010)]%
        {Eyben2010-zx}
\bibfield{author}{\bibinfo{person}{Florian Eyben}, \bibinfo{person}{Martin W{\"o}llmer}, {and} \bibinfo{person}{Bj{\"o}rn Schuller}.} \bibinfo{year}{2010}\natexlab{}.
\newblock \showarticletitle{{openSMILE} -- The Munich Versatile and Fast {Open-Source} Audio Feature Extractor}. In \bibinfo{booktitle}{\emph{Proceedings of the 9th {ACM} International Conference on Multimedia, {MM} 2010}}. \bibinfo{publisher}{unknown}, \bibinfo{pages}{1459--1462}.
\newblock


\bibitem[Funder and Sneed(1993)]%
        {Funder1993-xi}
\bibfield{author}{\bibinfo{person}{D~C Funder} {and} \bibinfo{person}{C~D Sneed}.} \bibinfo{year}{1993}\natexlab{}.
\newblock \showarticletitle{Behavioral manifestations of personality: an ecological approach to judgmental accuracy}.
\newblock \bibinfo{journal}{\emph{J. Pers. Soc. Psychol.}} \bibinfo{volume}{64}, \bibinfo{number}{3} (\bibinfo{date}{March} \bibinfo{year}{1993}), \bibinfo{pages}{479--490}.
\newblock


\bibitem[Grahe and Bernieri(1999)]%
        {Grahe1999-tm}
\bibfield{author}{\bibinfo{person}{Jon~E Grahe} {and} \bibinfo{person}{Frank~J Bernieri}.} \bibinfo{year}{1999}\natexlab{}.
\newblock \showarticletitle{The Importance of Nonverbal Cues in Judging Rapport}.
\newblock \bibinfo{journal}{\emph{J. Nonverbal Behav.}} \bibinfo{volume}{23}, \bibinfo{number}{4} (\bibinfo{date}{Dec.} \bibinfo{year}{1999}).
\newblock


\bibitem[Hagad et~al\mbox{.}(2011)]%
        {Hagad2011-od}
\bibfield{author}{\bibinfo{person}{Juan~Lorenzo Hagad}, \bibinfo{person}{Roberto Legaspi}, \bibinfo{person}{Masayuki Numao}, {and} \bibinfo{person}{Merlin Suarez}.} \bibinfo{year}{2011}\natexlab{}.
\newblock \showarticletitle{Predicting Levels of Rapport in Dyadic Interactions through Automatic Detection of Posture and Posture Congruence}. In \bibinfo{booktitle}{\emph{2011 {IEEE} Third International Conference on Privacy, Security, Risk and Trust and 2011 {IEEE} Third International Conference on Social Computing}}. \bibinfo{publisher}{IEEE}, \bibinfo{pages}{613--616}.
\newblock


\bibitem[Harrigan et~al\mbox{.}(1985)]%
        {Harrigan1985-cd}
\bibfield{author}{\bibinfo{person}{Jinni~A Harrigan}, \bibinfo{person}{Thomas~E Oxman}, {and} \bibinfo{person}{Robert Rosenthal}.} \bibinfo{year}{1985}\natexlab{}.
\newblock \showarticletitle{Rapport expressed through nonverbal behavior}.
\newblock \bibinfo{journal}{\emph{J. Nonverbal Behav.}} \bibinfo{volume}{9}, \bibinfo{number}{2} (\bibinfo{year}{1985}), \bibinfo{pages}{95--110}.
\newblock


\bibitem[Harris and Vazire(2016)]%
        {Harris2016-xv}
\bibfield{author}{\bibinfo{person}{Kelci Harris} {and} \bibinfo{person}{Simine Vazire}.} \bibinfo{year}{2016}\natexlab{}.
\newblock \showarticletitle{On friendship development and the Big Five personality traits}.
\newblock \bibinfo{journal}{\emph{Soc. Personal. Psychol. Compass}} \bibinfo{volume}{10}, \bibinfo{number}{11} (\bibinfo{date}{Nov.} \bibinfo{year}{2016}), \bibinfo{pages}{647--667}.
\newblock


\bibitem[Hayashi et~al\mbox{.}(2024)]%
        {Hayashi2024}
\bibfield{author}{\bibinfo{person}{Takato Hayashi}, \bibinfo{person}{Ryusei Kimura}, \bibinfo{person}{Ryo Ishii}, \bibinfo{person}{Fumio Nihei}, \bibinfo{person}{Atsushi Fukayama}, {and} \bibinfo{person}{Shogo Okada}.} \bibinfo{year}{2024}\natexlab{}.
\newblock \showarticletitle{Rapport Prediction Using Pairwise Learning in Dyadic Conversations Among Strangers and Among Friends}. In \bibinfo{booktitle}{\emph{Social Computing and Social Media}}, \bibfield{editor}{\bibinfo{person}{Adela Coman} {and} \bibinfo{person}{Simona Vasilache}} (Eds.). \bibinfo{publisher}{Springer Nature Switzerland}, \bibinfo{address}{Cham}, \bibinfo{pages}{17--28}.
\newblock
\showISBNx{978-3-031-61312-8}


\bibitem[Hayashi et~al\mbox{.}(2023)]%
        {Hayashi2023-bp}
\bibfield{author}{\bibinfo{person}{Takato Hayashi}, \bibinfo{person}{Candy~Olivia Mawalim}, \bibinfo{person}{Ryo Ishii}, \bibinfo{person}{Akira Morikawa}, \bibinfo{person}{Atsushi Fukayama}, \bibinfo{person}{Takao Nakamura}, {and} \bibinfo{person}{Shogo Okada}.} \bibinfo{year}{2023}\natexlab{}.
\newblock \showarticletitle{A Ranking Model for Evaluation of Conversation Partners Based on Rapport Levels}.
\newblock \bibinfo{journal}{\emph{IEEE Access}}  \bibinfo{volume}{11} (\bibinfo{year}{2023}), \bibinfo{pages}{73024--73035}.
\newblock


\bibitem[Huang et~al\mbox{.}(2011)]%
        {Huang2011-qc}
\bibfield{author}{\bibinfo{person}{Lixing Huang}, \bibinfo{person}{Louis-Philippe Morency}, {and} \bibinfo{person}{Jonathan Gratch}.} \bibinfo{year}{2011}\natexlab{}.
\newblock \showarticletitle{Virtual Rapport 2.0}. In \bibinfo{booktitle}{\emph{Intelligent Virtual Agents}}. \bibinfo{publisher}{Springer Berlin Heidelberg}, \bibinfo{pages}{68--79}.
\newblock


\bibitem[Joel et~al\mbox{.}(2017)]%
        {Joel2017-ne}
\bibfield{author}{\bibinfo{person}{Samantha Joel}, \bibinfo{person}{Paul~W Eastwick}, {and} \bibinfo{person}{Eli~J Finkel}.} \bibinfo{year}{2017}\natexlab{}.
\newblock \showarticletitle{Is Romantic Desire Predictable? Machine Learning Applied to Initial Romantic Attraction}.
\newblock \bibinfo{journal}{\emph{Psychol. Sci.}} \bibinfo{volume}{28}, \bibinfo{number}{10} (\bibinfo{date}{Oct.} \bibinfo{year}{2017}), \bibinfo{pages}{1478--1489}.
\newblock


\bibitem[Kenny(2019)]%
        {Kenny2019-ji}
\bibfield{author}{\bibinfo{person}{David~A Kenny}.} \bibinfo{year}{2019}\natexlab{}.
\newblock \bibinfo{booktitle}{\emph{Interpersonal Perception: The Foundation of Social Relationships}}.
\newblock \bibinfo{publisher}{Guilford Publications}.
\newblock


\bibitem[Kenny and La~Voie(1984)]%
        {Kenny1984-su}
\bibfield{author}{\bibinfo{person}{David~A Kenny} {and} \bibinfo{person}{Lawrence La~Voie}.} \bibinfo{year}{1984}\natexlab{}.
\newblock \showarticletitle{The Social Relations Model}.
\newblock In \bibinfo{booktitle}{\emph{Advances in Experimental Social Psychology}}, \bibfield{editor}{\bibinfo{person}{Leonard Berkowitz}} (Ed.). Vol.~\bibinfo{volume}{18}. \bibinfo{publisher}{Academic Press}, \bibinfo{pages}{141--182}.
\newblock


\bibitem[Kimura et~al\mbox{.}(2005)]%
        {Kimura2005-dw}
\bibfield{author}{\bibinfo{person}{Masanori Kimura}, \bibinfo{person}{Masao Yogo}, {and} \bibinfo{person}{Ikuo Daibo}.} \bibinfo{year}{2005}\natexlab{}.
\newblock \showarticletitle{Expressivity halo effect in the conversation about emotional episodes}.
\newblock \bibinfo{journal}{\emph{THE JAPANESE JOURNAL OF RESEARCH ON EMOTIONS}} \bibinfo{volume}{12}, \bibinfo{number}{1} (\bibinfo{year}{2005}), \bibinfo{pages}{12--23}.
\newblock


\bibitem[Lin(1989)]%
        {Lin1989ACC}
\bibfield{author}{\bibinfo{person}{Lawrence I-Kuei Lin}.} \bibinfo{year}{1989}\natexlab{}.
\newblock \showarticletitle{A concordance correlation coefficient to evaluate reproducibility.}
\newblock \bibinfo{journal}{\emph{Biometrics}}  \bibinfo{volume}{45 1} (\bibinfo{year}{1989}), \bibinfo{pages}{255--68}.
\newblock
\urldef\tempurl%
\url{https://api.semanticscholar.org/CorpusID:32656801}
\showURL{%
\tempurl}


\bibitem[Madaio et~al\mbox{.}(2017)]%
        {Madaio2017-xr}
\bibfield{author}{\bibinfo{person}{Michael Madaio}, \bibinfo{person}{Rae Lasko}, \bibinfo{person}{Amy Ogan}, {and} \bibinfo{person}{Justine Cassell}.} \bibinfo{year}{2017}\natexlab{}.
\newblock \showarticletitle{Using Temporal Association Rule Mining to Predict Dyadic Rapport in Peer Tutoring}. In \bibinfo{booktitle}{\emph{International Conference of Educational Data Mining}}. \bibinfo{publisher}{unknown}.
\newblock


\bibitem[Martínez et~al\mbox{.}(2014)]%
        {Martínez2014}
\bibfield{author}{\bibinfo{person}{Héctor~P. Martínez}, \bibinfo{person}{Georgios~N. Yannakakis}, {and} \bibinfo{person}{John Hallam}.} \bibinfo{year}{2014}\natexlab{}.
\newblock \showarticletitle{Don’t Classify Ratings of Affect; Rank Them!}
\newblock \bibinfo{journal}{\emph{IEEE Transactions on Affective Computing}} \bibinfo{volume}{5}, \bibinfo{number}{3} (\bibinfo{year}{2014}), \bibinfo{pages}{314--326}.
\newblock
\urldef\tempurl%
\url{https://doi.org/10.1109/TAFFC.2014.2352268}
\showDOI{\tempurl}


\bibitem[Mawalim et~al\mbox{.}(2023)]%
        {Mawalim2023-sq}
\bibfield{author}{\bibinfo{person}{Candy~Olivia Mawalim}, \bibinfo{person}{Shogo Okada}, \bibinfo{person}{Yukiko~I Nakano}, {and} \bibinfo{person}{Masashi Unoki}.} \bibinfo{year}{2023}\natexlab{}.
\newblock \showarticletitle{Personality trait estimation in group discussions using multimodal analysis and speaker embedding}.
\newblock \bibinfo{journal}{\emph{J. Multimodal User Interfaces}} \bibinfo{volume}{17}, \bibinfo{number}{2} (\bibinfo{date}{June} \bibinfo{year}{2023}), \bibinfo{pages}{47--63}.
\newblock


\bibitem[Miles et~al\mbox{.}(2009)]%
        {MILES2009585}
\bibfield{author}{\bibinfo{person}{Lynden~K. Miles}, \bibinfo{person}{Louise~K. Nind}, {and} \bibinfo{person}{C.~Neil Macrae}.} \bibinfo{year}{2009}\natexlab{}.
\newblock \showarticletitle{The rhythm of rapport: Interpersonal synchrony and social perception}.
\newblock \bibinfo{journal}{\emph{Journal of Experimental Social Psychology}} \bibinfo{volume}{45}, \bibinfo{number}{3} (\bibinfo{year}{2009}), \bibinfo{pages}{585--589}.
\newblock
\showISSN{0022-1031}
\urldef\tempurl%
\url{https://doi.org/10.1016/j.jesp.2009.02.002}
\showDOI{\tempurl}


\bibitem[M{\"u}ller et~al\mbox{.}(2018)]%
        {Muller2018-ax}
\bibfield{author}{\bibinfo{person}{Philipp M{\"u}ller}, \bibinfo{person}{Michael~Xuelin Huang}, {and} \bibinfo{person}{Andreas Bulling}.} \bibinfo{year}{2018}\natexlab{}.
\newblock \showarticletitle{Detecting Low Rapport During Natural Interactions in Small Groups from {Non-Verbal} Behaviour}. In \bibinfo{booktitle}{\emph{Proceedings of the 23rd International Conference on Intelligent User Interfaces}} (<conf-loc>, <city>Tokyo</city>, <country>Japan</country>, </conf-loc>) \emph{(\bibinfo{series}{IUI '18})}. \bibinfo{publisher}{Association for Computing Machinery}, \bibinfo{address}{New York, NY, USA}, \bibinfo{pages}{153--164}.
\newblock


\bibitem[Poria et~al\mbox{.}(2017)]%
        {Poria2017-nf}
\bibfield{author}{\bibinfo{person}{Soujanya Poria}, \bibinfo{person}{Erik Cambria}, \bibinfo{person}{Devamanyu Hazarika}, \bibinfo{person}{Navonil Majumder}, \bibinfo{person}{Amir Zadeh}, {and} \bibinfo{person}{Louis-Philippe Morency}.} \bibinfo{year}{2017}\natexlab{}.
\newblock \showarticletitle{{Context-Dependent} Sentiment Analysis in {User-Generated} Videos}. In \bibinfo{booktitle}{\emph{Proceedings of the 55th Annual Meeting of the Association for Computational Linguistics (Volume 1: Long Papers)}}, \bibfield{editor}{\bibinfo{person}{Regina Barzilay} {and} \bibinfo{person}{Min-Yen Kan}} (Eds.). \bibinfo{publisher}{Association for Computational Linguistics}, \bibinfo{address}{Vancouver, Canada}, \bibinfo{pages}{873--883}.
\newblock


\bibitem[Sharma et~al\mbox{.}(2021)]%
        {Sharma2021-mi}
\bibfield{author}{\bibinfo{person}{Srijan Sharma}, \bibinfo{person}{Kantha~Girish Gangadhara}, \bibinfo{person}{Fei Xu}, \bibinfo{person}{Anne~Solbu Slowe}, \bibinfo{person}{Mark~G Frank}, {and} \bibinfo{person}{Ifeoma Nwogu}.} \bibinfo{year}{2021}\natexlab{}.
\newblock \showarticletitle{Coupled Systems for Modeling Rapport Between Interlocutors}. In \bibinfo{booktitle}{\emph{2021 16th {IEEE} International Conference on Automatic Face and Gesture Recognition ({FG} 2021)}}. \bibinfo{publisher}{IEEE}, \bibinfo{pages}{1--8}.
\newblock


\bibitem[Sinha and Cassell(2015)]%
        {Sinha2015-lg}
\bibfield{author}{\bibinfo{person}{Tanmay Sinha} {and} \bibinfo{person}{Justine Cassell}.} \bibinfo{year}{2015}\natexlab{}.
\newblock \showarticletitle{We Click, We Align, We Learn: Impact of Influence and Convergence Processes on Student Learning and Rapport Building}. In \bibinfo{booktitle}{\emph{Proceedings of the 1st Workshop on Modeling {INTERPERsonal} {SynchrONy} And infLuence}} (Seattle, Washington, USA) \emph{(\bibinfo{series}{INTERPERSONAL '15})}. \bibinfo{publisher}{Association for Computing Machinery}, \bibinfo{address}{New York, NY, USA}, \bibinfo{pages}{13--20}.
\newblock


\bibitem[Tickle-Degnen and Rosenthal(1990)]%
        {Tickle-Degnen1990-jp}
\bibfield{author}{\bibinfo{person}{Linda Tickle-Degnen} {and} \bibinfo{person}{Robert Rosenthal}.} \bibinfo{year}{1990}\natexlab{}.
\newblock \showarticletitle{The Nature of Rapport and Its Nonverbal Correlates}.
\newblock \bibinfo{journal}{\emph{Psychol. Inq.}} \bibinfo{volume}{1}, \bibinfo{number}{4} (\bibinfo{date}{Oct.} \bibinfo{year}{1990}), \bibinfo{pages}{285--293}.
\newblock


\bibitem[Wada(1996)]%
        {Wada1996-pk}
\bibfield{author}{\bibinfo{person}{Sayuri Wada}.} \bibinfo{year}{1996}\natexlab{}.
\newblock \showarticletitle{Construction of the Big Five Scales of personality trait terms and concurrent validity with {NPI}}.
\newblock \bibinfo{journal}{\emph{Shinrigaku Kenkyu}} \bibinfo{volume}{67}, \bibinfo{number}{1} (\bibinfo{date}{April} \bibinfo{year}{1996}), \bibinfo{pages}{61--67}.
\newblock


\bibitem[Wang and Gratch(2009)]%
        {Wang2009-mh}
\bibfield{author}{\bibinfo{person}{Ning Wang} {and} \bibinfo{person}{Jonathan Gratch}.} \bibinfo{year}{2009}\natexlab{}.
\newblock \showarticletitle{Rapport and facial expression}. In \bibinfo{booktitle}{\emph{2009 3rd International Conference on Affective Computing and Intelligent Interaction and Workshops}}. \bibinfo{publisher}{IEEE}, \bibinfo{pages}{1--6}.
\newblock


\bibitem[Wood et~al\mbox{.}(2010)]%
        {Wood2010-sa}
\bibfield{author}{\bibinfo{person}{Dustin Wood}, \bibinfo{person}{Peter Harms}, {and} \bibinfo{person}{Simine Vazire}.} \bibinfo{year}{2010}\natexlab{}.
\newblock \showarticletitle{Perceiver effects as projective tests: what your perceptions of others say about you}.
\newblock \bibinfo{journal}{\emph{J. Pers. Soc. Psychol.}} \bibinfo{volume}{99}, \bibinfo{number}{1} (\bibinfo{date}{July} \bibinfo{year}{2010}), \bibinfo{pages}{174--190}.
\newblock


\bibitem[Zhao et~al\mbox{.}(2016)]%
        {Zhao2016-dk}
\bibfield{author}{\bibinfo{person}{Ran Zhao}, \bibinfo{person}{Tanmay Sinha}, \bibinfo{person}{Alan~W Black}, {and} \bibinfo{person}{Justine Cassell}.} \bibinfo{year}{2016}\natexlab{}.
\newblock \showarticletitle{Socially-aware virtual agents: Automatically assessing dyadic rapport from temporal patterns of behavior}.
\newblock In \bibinfo{booktitle}{\emph{Intelligent Virtual Agents}}. \bibinfo{publisher}{Springer International Publishing}, \bibinfo{address}{Cham}, \bibinfo{pages}{218--233}.
\newblock


\end{thebibliography}










\end{document}